# Unusual gas sensor response and semiconductor-to-insulator transition in $WO_{3-x}$ nanostructures : The role of oxygen vacancy

K. Ganesan [a,b,1] and P.K. Ajikumar [a]

[a] *Materials Science Group, Indira Gandhi Centre for Atomic Research, Kalpakkam - 603102, India.*
[b] *Homi Bhabha National Institute, Training School Complex, Anushaktinagar, Mumbai- 400094, India*

## Abstract

$WO_{3-x}$ thinfilms featuring petal-like and lamella-like nanostructures are grown under controlled oxygen partial pressures using hot filament chemical vapor deposition. These synthesized $WO_{3-x}$ nanostructures exhibit monoclinic structure and contain a significant amount of oxygen vacancies ($V_O$) as confirmed by X-ray diffraction and Raman spectroscopy, respectively. These $WO_{3-x}$ nanostructures demonstrate sensor response to both $NH_3$ and $NO_2$ gases even at room temperature. However, the sensor response varies with temperature and analyte gas type. For $NH_3$, the sensors exhibit an increase in resistance behaving like a p-type semiconductor at temperatures below 150 $^0C$ while the resistance decreases at higher temperatures, resembling n-type semiconductor behavior. On the other hand, below 150 $^0C$, these sensors display n-type behavior towards $NO_2$ but act like p-type semiconductor at higher temperatures. Further temperature dependent transport studies were performed in these $WO_{3-x}$ nanostructures in the temperature range of 25 – 300 $^oC$, after inducing additional $V_O$ in the films through annealing under $CH_4$. The petal-like $WO_{3-x}$ nanostructures display an abrupt change in resistance, indicating insulator-to-semiconductor and semiconductor-to-insulator transitions during heating and cooling cycles respectively, in the temperature range of 100 - 212 $^0C$. In lamella-like $WO_{3-x}$ nanostructures, the resistance is flipped from semiconductor-to-insulator at 300 $^0C$ and remains insulating state when cooled down to 30 $^0C$. The abnormal gas sensing behavior and insulator - semiconductor transition is discussed in terms of $V_O$ in $WO_{3-x}$ nanostructures.

---

[1] Corresponding author. Email : kganesan@igcar.gov.in ( K. Ganesan)

# 1. Introduction

The metal oxide based resistive sensors have attracted much attention for detection of chemical gases down to ppb level because of their fast response, reproducibility, compactness, cost effectiveness and portability [1–4]. The semiconducting oxides such as $SnO_2$, ZnO, $WO_3$, $TiO_2$, and $In_2O_3$ were traditionally used for chemical sensor applications with high sensitivity at relatively lower operating temperatures. Further, these metal oxides with different microstructures, morphologies, oxygen stoichiometry and also, metal decorated and hybrid metal oxide heterostructures were investigated for improving the sensitivity and selectivity of the sensors towards various gases [2,3,5–7]. In recent years, the electronic nose which uses an array of sensors that may be of different materials and operating temperatures, had become a common tool in the identification and quantification of pollutant chemicals [8]. Apart from conventional oxides, two dimensional layered structures such as graphene, metal chalcogenides and Mxenes are also becoming popular topic of research for sensing applications [9].

Among the various metal oxide semiconductors, $WO_3$ is one of the most studied materials for potential applications in chemical sensors, electrochromic devices and catalysis [2,3,6,7,10–27]. Intensive studies were also carried out on $WO_3$ for the detection of various chemical gases, preferably for detection of $NO_2$ and $NH_3$ gases and volatile organic compounds because of its excellent chemical and physical properties [1–3,7,13,15,20,21,25,26]. $WO_3$ with different morphologies such as quantum dots (0D), nanowires (1D), nanosheets (2D) and porous nanostructures (3D) were studied for sensor applications [23]. Depending upon the requirement, $WO_3$ materials were prepared in thin film or powder form using various techniques like chemical vapour deposition (CVD), sputtering, sol-gel, thermal evaporation, hydrothermal methods and pulsed laser deposition (PLD) [10,24–27]. Hot filament CVD has the advantages for the preparation of porous 3D-like structures with controlled morphology and oxygen stoichiometry. The oxygen off-stoichiometry is one of the deterministic parameters for controlling the sensitivity and operation temperature of metal oxide based gas sensors. Further, the electrical characteristics of $WO_3$ can be vastly tuned by controlling the oxygen stoichiometry [11–15]. Moreover, Wang et. al. [28] had predicted the strong dependence of $WO_{3-x}$ electronic properties including semiconductor-to-metal transition (SMT) on the oxygen vacancies ($V_O$) using density functional theory. In addition, a few researchers had also observed metal to insulator transition (MIT) on the $WO_3$ through electrochemical gating. However, there is no report exists in literature on the observation of SMT or MIT transition through temperature dependent transport measurements in $WO_3$.

The $WO_3$ sensors had displayed excellent sensitivity for different gases at higher temperatures (> 150 $^0$C). However, the studies on sensor response at lower temperatures especially at room temperature were limited. Also, the sensor response of almost all the reported studies reveals that the $WO_3$ sensors behave as conventional n-type oxide semiconductor whose resistance decreases (increases) for reducing (oxidizing) gases, irrespective of the operating temperature . On the other hand, only a very few reports suggest an anomalous sensor behavior of $WO_{3-x}$ devices whose resistance increases at low temperatures for reducing gases, as if p-type semiconductor while it behaves as normal n-type semiconductor at higher temperatures. Kim et al [11] and Zhao et al [15] had reported an unusual response of $WO_{3-x}$ sensors and they had attributed to the oxygen off-stoichiometry. Li et al [13], had observed a similar anomalous behavior for ammonia and ethanol (both reducing) gases for $WO_3$-$SnO_2$ based gas sensors at different temperatures and they had attributed to the presence of adsorbed moisture at the surface of the sensor material along with adsorbed oxygen. Though different mechanisms were proposed for the anomalous sensing behavior of $WO_3$ nanostructures, there is no clear understanding on the materials aspects, why do only a certain type of $WO_{3-x}$ nanostructures display the unusual p to n carrier type conversion with operating temperature. These two above mentioned aspects, SMT or insulator – semiconductor transition (IST) and unusual sensor response, motivate us to probe further into the $WO_{3-x}$ nanostructures to understand the underlying mechanism. Moreover, tuning the oxygen vacancy concentration offers excellent opportunities to enhance the electrical, optical, and catalytic properties of $WO_{3-x}$ nanostructures. This improvement paves the way for advanced applications in electrochromic devices, photocatalysis, and electronic circuits. Additionally, $WO_3$ is a stable, environmentally friendly, non-toxic, and cost-effective material that can be seamlessly integrated into existing device technologies.

In the present study, we have prepared the oxygen sub-stoichiometric $WO_{3-x}$ nanostructures with two different morphologies. Then, the deposited $WO_{3-x}$ nanostructures were characterized by scanning electron microscopy (SEM), energy dispersive X-ray analysis (EDX), X-ray diffraction (XRD) and Raman spectroscopy for their morphology and crystal structure. The resistive gas sensing studies were carried out for both reducing ($NH_3$) and oxidizing ($NO_2$) gases. The sensor devices exhibit conversion from p to n-type or vice versa as a function of temperature for both reducing and oxidizing gases, respectively. Further, the $V_O$ on these WO$_{3-x}$ nanostructures were increased by annealing under reducing atmosphere of CH$_4$ at 300 $^0$C and probed the correlation between electrical properties and $V_O$ through temperature dependent resistance measurements. Based on the observations, the abnormal sensing behavior and IST in WO$_{3-x}$ nanostructures are attributed to the oxygen vacancies in the lattice.

## 2. Experimental

WO$_{3-x}$ nanostructures were deposited directly on 0.5 mm thick alumina substrate having the size of 25 x 5 mm$^2$ with pre-patterned inter-digitated platinum electrodes, using hot filament CVD technique and the details can be found elsewhere [29]. This inter-digitated electrode has 20 fingers with spacing between the electrodes of 200 μm. A high pure (99.99 %) tungsten wire was heated with a DC power supply in controlled oxygen flow of 1 sccm for preparing the nanostructures. The feedstock oxygen reacts aggressively with hot *W* filament surface which readily oxidizes into WO$_{3-x}$. Since the vapor pressure of WO$_3$ is high, it evaporates concurrently and get deposited on the substrate which is kept at 800 $^0$C. Exposure time, oxygen partial pressure and DC power were varied to optimize the thickness and morphology of the coatings. Typical growth duration was 30 s. WO$_{3-x}$ nanostructures with two different morphologies viz. petal-like and lamella-like nanostructures were grown at the filament power of 100 and 150 W while other experimental parameters were kept constant and these structures are labeled as S1 and S2, respectively.

The crystal structure of the WO$_{3-x}$ nanostructures were characterized by glancing incident X-ray diffraction (XRD) (Inel, equinox 2000) with copper Kα radiation (λ=1.54 A). FESEM (Supra55, Carl Zeiss) was used for surface morphological analysis. The elemental composition of the WO$_{3-x}$ nanostructures that were grown simultaneously on Si substrates were also analyzed using EDX analysis (Supra55, Carl Zeiss). EDX measurements were performed over an area of ~ 40 × 40 μm² at a minimum of four different locations to ensure compositional homogeneity in the samples. Raman spectrometer (Invia, Renishaw) with laser wavelength of 532 nm in back scattering geometry was used for analyzing vibrational properties of the nanostructures.

The dynamic gas sensor measurements on both the samples were carried out simultaneously using a computer interfaced data acquisition system. The sensor devices were placed on a flat heater inside a vacuum chamber with electrical leads through vacuum feed-through [30]. The analyte gases NH$_3$ and NO$_2$ which were diluted down to 1000 and 100 ppm in ultra-high pure $N_2$ respectively, were used for the study. For sensor measurements, these pre-diluted analyte gases were further mixed with synthetic air (90 % of $N_2$ and 10 % of $O_2$) and were admitted to the evacuated chamber using mass flow controllers. The total flow of the input gases were kept at constant of 100 sccm. The chamber background pressure was maintained at 500 mbar using throttling valve throughout the sensor measurements. The sensor measurements were carried out at the temperatures, 25, 50, 100, 150, 200, and 300 $^0$C at concentrations of 100, 500 and 1000 ppm of $NH_3$ and at 5, 10 and 15 ppm for $NO_2$. The analyte gas flow ON and OFF time were maintained for 300 s, respectively.

The temperature dependent resistance of the samples S1 and S2 that were used for sensor study were measured simultaneously using two channel source and measure unit by two probe method in the temperature range of 30 – 300 $^0$C. After completing the first set of measurements, the samples S1 and S2 were annealed under reducing atmosphere of 1000 ppm of $CH_4$ diluted in $N_2$ at 300 $^0$C under 650 mbar for 2 h, in order to create additional $V_O$ in the $WO_{3-x}$ lattice and these samples are labeled as S1_CH$_4$ and S2_CH$_4$, respectively. In addition, these samples were also exposed to ambient atmosphere for 12 h at 25 $^0$C and then, transport measurements were repeated for third time. These samples are labeled as S1_CH$_4$_Air and S2_CH$_4$_Air. All the electrical resistance measurements were performed under vacuum of ~ 8 x 10$^{-3}$ mbar. Typical heating and cooling rate were kept at 2.5 $^0$C/min and also, the resistance measurements were carried out during heating and cooling cycles.

## 3. Results
### 3.1. Surface morphology and structural analysis

Figures 1a and 1b show the typical FESEM morphology of the samples S1 and S2 and their magnified images are also given in Figs. 1c and 1d, respectively. The morphology of the sample S1 consists of nearly oblong shaped grains with petal-like structures having ~ 200 nm length and sharp edges (Fig. 1c). On the other hand, the sample S2 has a lamella-like structure with relatively thicker wall nanostructures as compared to that of sample S1. Further, both these $WO_{3-x}$ nanostructures are highly porous in nature with inter connected networks in three dimensional space as can be evidenced from the micrographs in Figs. 1c and 1d. However, qualitative FESEM analysis indicates that sample S1 has a higher surface-to-volume ratio compared to sample S2. The thickness of the coatings is about 400 and 700 nm as measured from the cross sectional FESEM micrographs of the samples S1 and S2, respectively.

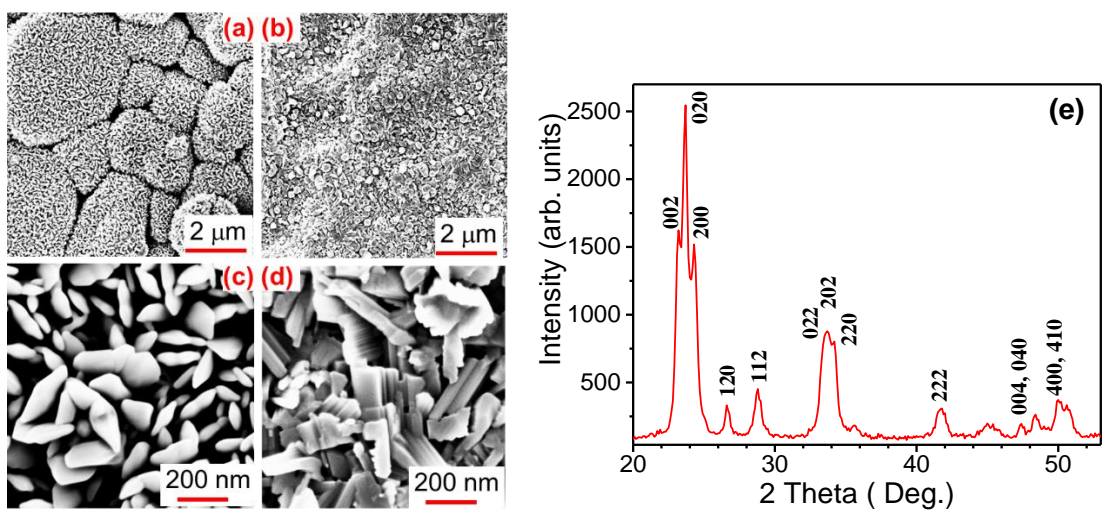

Fig. 1. Scanning electron micrographs of WO$_{3-x}$ nanostructures grown at DC power of (a) 100 W and (b) 150 W and a magnified part of these micrographs are given in (c) and (d), respectively. (e) Typical X-ray diffraction pattern of synthesized WO$_{3-x}$ nanostructures.

Based on EDX analysis, the elemental compositions of the WO$_{3-x}$ nanostructures were found to be WO$_{2.43 \pm 0.01}$ and WO$_{2.56 \pm 0.01}$ for petal-like (S1) and lamella-like (S2) nanostructures, respectively. This result indicates that the oxygen vacancy concentration (x) in WO$_{3-x}$ is approximately 0.57 ± 0.01 for petal-like structures and 0.44 ± 0.01 for lamella-like structures. We note here that the sensitivity of EDX analysis for oxygen is around 2 atomic %, which may introduce minor errors in the accuracy of the oxygen vacancy concentration. However, this error is expected to be consistent across both samples. Importantly, the EDX analysis clearly demonstrates a higher oxygen vacancy concentration in sample S1 compared to sample S2. Further, a typical GIXRD pattern of these WO$_3$ nanostructures is shown in Fig. 1e. The peak positions of the spectrum matches well with monoclinic phase of WO$_3$ (JCPDS 43-1035). There are no additional peaks in the diffraction pattern indicating the phase purity of WO$_{3-x}$ nanostructures. A little peak broadening and overlapping of the XRD peaks indicate the nanocrystalline nature of synthesized WO$_{3-x}$ films with reasonable crystalline quality.

### 3.2. Raman spectroscopy

Figure 2 shows the typical Raman spectra of the grown petal-like and lamella-like WO$_{3-x}$ nanostructures in the range from 100 to 1200 cm$^{-1}$. The Raman spectra of these nanostructures appear nearly identical and display sharp Raman bands which indicate the good crystalline nature of the WO$_{3-x}$ nanostructures. The most prominent modes at ~ 804 and 715 cm$^{-1}$ are attributed to the stretching and bending modes of W-O bonds, and deformation of O-W-O bonds respectively. These vibrational modes also confirm the monoclinic phase of the synthesized WO$_{3-x}$ nanostructures [16]. Also, there are several low frequency Raman bands at ~ 130, 180, 217, 270, 324 and 435 cm$^{-1}$ which are associated with oxygen vacancy (V$_O$) defects in the WO$_3$ nanostructures. The Raman band at 180 cm$^{-1}$ is assigned to the vibration of W$^{5+}$-W$^{5+}$ stretching with an oxygen vacancy [17]. The modes at 270 and 324 cm$^{-1}$ are attributed to bending motions of O-W$^{5+}$-O bonds within the octahedra of crystalline WO$_3$ [17]. Overall, supported by EDX analysis, Raman spectroscopic study clearly indicates that the synthesized WO$_{3-x}$ nanostructures have a significant amount of V$_O$ which manifest as Raman modes below 500 cm$^{-1}$, while retaining the reasonably good crystalline nature.

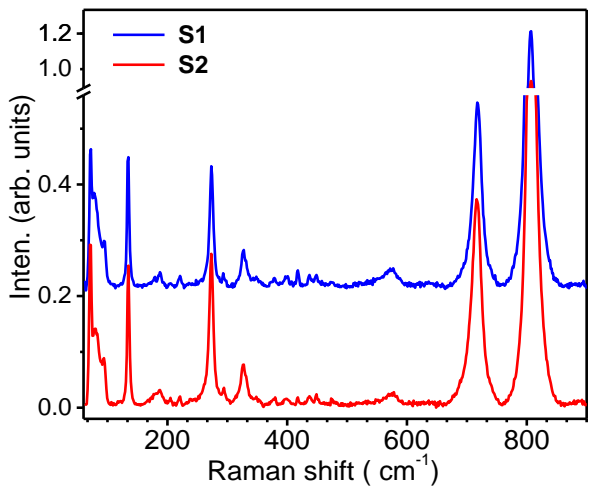

Fig. 2. Raman spectra of WO$_{3-x}$ nanostructures grown at filament power of 100 (S1) and 150 W (S2). The spectra are normalized with high intensity band and stacked vertical for clarity.

### 3.3. Gas sensor measurement

Figures 3a and 3b show the gas sensor response of the samples S1 and S2 that are exposed to NH$_3$ at different analyte concentrations in the range of 100 – 1000 ppm and operating temperatures of 25 – 300 $^0$C. The sensor response (R) is calculated based on the equation,

$$R\ (\%) = \frac{(R_g - R_i) \times 100}{R_i} \quad \ldots\ldots\ldots\ldots\ldots\ldots (1)$$

where, $R_i$ and $R_g$ represent the resistance of the sensor device initial and after exposing to analyte gas, respectively. Since the intrinsic WO$_3$ is a well-known n-type semiconductor due to V$_O$, the resistance of the device was expected to decrease for exposing to NH$_3$ since it donates electron to WO$_3$ structures. In contrast, the resistance of both the sensor devices is increased with the exposure of NH$_3$ at temperatures 25, 50 and 100 $^0$C. At 150$^0$C, the resistance of the sensor S1 increases rapidly and subsequently, a monotonic decrease in resistance is observed. On the other hand, the sensor S2 displays sudden increase in resistance initially, subsequently, the resistance decreases gradually. At 200 and 300 $^0$C, the resistance of the both the sensor devices decreases continuously upon exposure to NH$_3$. Also, the resistance falling slope of the devices is steeper at 300 $^0$C as compared to 200 $^0$C, as can be seen from Figs. 3a and 3b. Though these two sensors behave in same manner in the studied temperature range, the response of sensor S1 is much higher as compared to the sensor S2. Since both sensors have high V$_O$ concentration, with x = 0.57 ± 0.01 for S1 and x = 0.44 ± 0.01 for S2, they exhibit similar behavior within the studied temperature range. However, the sensor S1 demonstrates a significantly higher response compared to sensor S2, as its surface to volume ratio is greater.

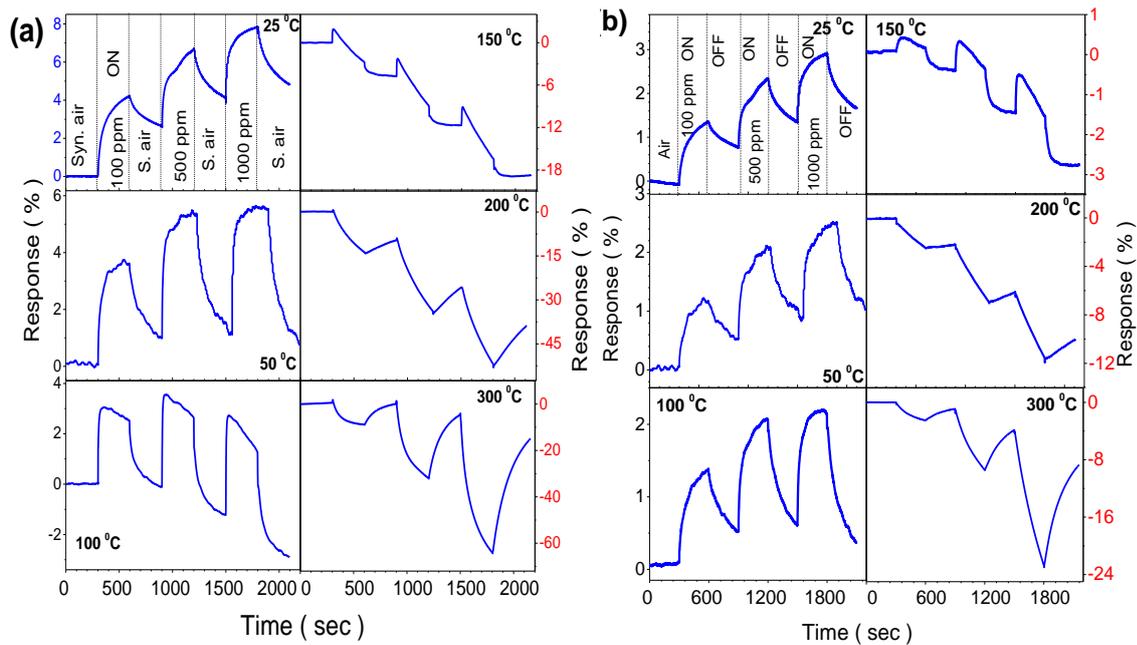

Fig. 3. The sensor response of WO$_3$ nanostructures at different concentrations of NH$_3$ and operating temperatures for (a) sample S1 (b) sample S2.

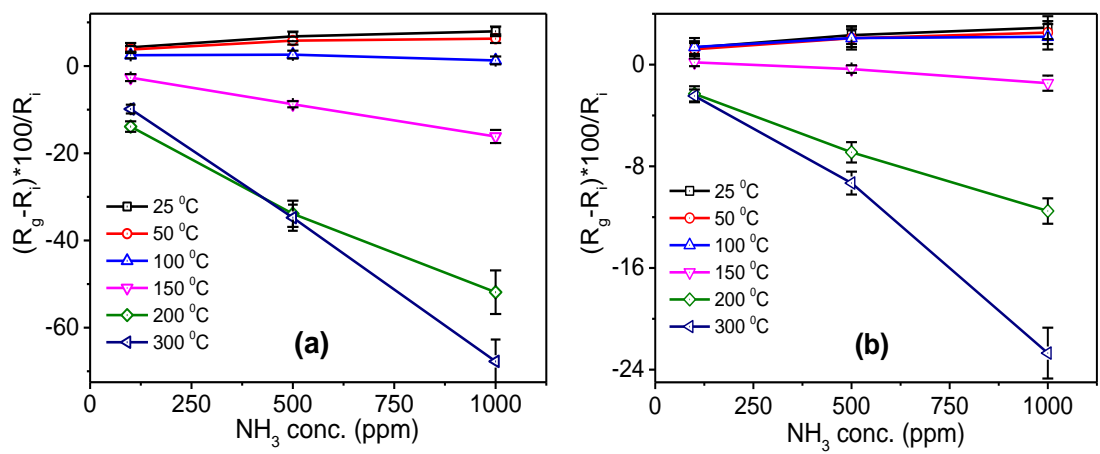

Fig. 4. The concentration dependent sensor response of $WO_3$ at different temperatures for the analyte gas, $NH_3$ of the samples (a) S1 and (b) S2

Figures 4a and 4b show the response of the $WO_3$ sensor devices S1 and S2 respectively, as a function of $NH_3$ concentration at different temperatures. Here, the response is calculated based on the equation 1 wherein the value of $R_g$ is taken after exposing the device with analyte gas for 300 s. Both the sensor devices exhibit positive response for $NH_3$ at temperatures $\leq 100\ ^0C$ while it shows negative response at higher temperatures, $\geq 150\ ^0C$. Further, these sensors show a nearly linear response as a function of analyte concentration in the range of 100 – 1000 ppm for $NH_3$. The sensor response changes from positive to negative value around $\sim 150\ ^0C$ and it is clearly evident from Figs. 4a and 4b. These results indicate that two competing mechanisms are contributing to the electronic conduction processes in the studied temperature range of 23 – 300 $^0C$. Moreover, the sensitivity which is defined as response of a device for a given concentration of analyte gas is much higher for the device S1 as compared to S2.

As similar to $NH_3$ test, the sensor measurements were also repeated for $NO_2$ under identical conditions except the concentration which was kept at 5, 10 and 15 ppm diluted in $N_2$. Figs. 5a and 5b show the sensor response of sample S1 and S2, respectively when exposed to $NO_2$ gas at different analyte concentrations and operating temperatures. At 25, 50 and 100 $^0C$, the resistance of the sensor device S1 increases indicating n-type behavior of $WO_3$ nanostructures. At 150 $^0C$, the response was a sudden increase followed by monotonic decrease in resistance, although the overall change in resistivity was small. At 200 $^0C$, an instant increase and an immediate decrease followed by continuous decrease in resistance of the device were observed. The same trend is also observed at 300 $^0C$ with much higher rate of change in resistance. The sudden jump in increase of resistance and subsequent gradual decrease of resistance at $\geq 150^0$C, clearly reveals more than one competing mechanisms in action upon $NO_2$ exposures. . Furthermore, the response of the sensor device S2 also follows a similar trend to the device S1, however, the instant increase and subsequent decrease in resistance immediately after exposure to $NO_2$ is not observed at high temperatures $> 150\ ^0$C. Moreover, the sensitivity of the device S1 is slightly higher than that of S2 for $NO_2$.

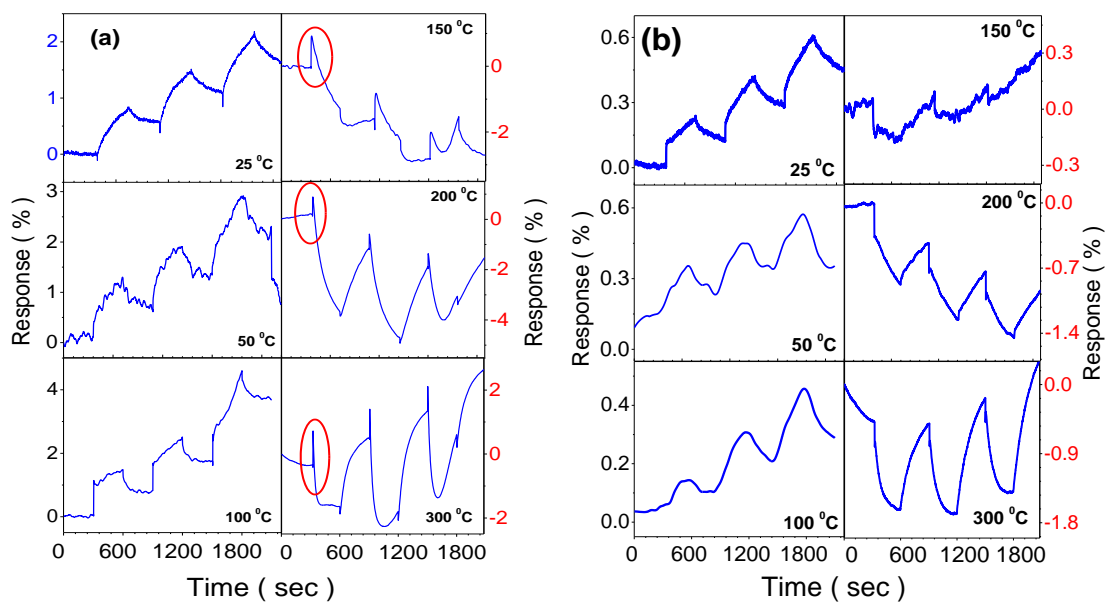

Fig. 5. The sensor response of $WO_{3-x}$ nanostructures at different concentrations and operating temperatures for analyte gas $NO_2$ , (a) S1 and (b) S2

A comparative gas sensor response of our samples with literature data is shown in Table 1. Though the response of the grown $WO_{3-x}$ nanostructures is not so high, it displays a reasonable response at room temperature. Further, the $WO_{3-x}$ nanostructures exhibit excellent sensor stability and repeatability even after two months. Additionally, the response of sensor devices fabricated from different batches of $WO_{3−x}$ nanostructures shows minimal variation. However, the base resistance of the devices exhibits slight changes between batches.

Table 1. A comparative gas sensor performance of the WO$_{3-x}$ nanostructures with literature data

| Material | p to n / n to p type conversion & temperature | Analyte gas | concentration | Lowest response temp. (°C) | Response (R$_g$/R$_i$ or R$_i$/R$_g$) | Reference |
|---|---|---|---|---|---|---|
| WO$_3$-SnO$_2$ | Yes, 95 °C | NH$_3$ | 0.2 vol % | 25 | 1.14 – 1.72 | [13] |
| W$_{18}$O$_{49}$ nanowire | Yes, 25 °C with NH$_3$ concentration | NH$_3$ | 0.1 – 400 ppm | 25 | - | [15] |
| Flower like WO$_3$ | No | NH$_3$ | 10 ppm | 120 | ~ 2 | [31] |
| WO$_3$ nanorod | No | NH$_3$ | 100 ppm | 50 | ~ 3 | [32] |
| Sb-doped WO$_3$ | No | NH$_3$ | 100 ppm | 20 | ~ 2 | [33] |
| WO$_3$ | Yes, 130 °C | NO$_2$ | < 450 ppb | 95 | 1 - 28 | [12] |
| WO3 | Yes, 250 °C | NO$_2$ | 100 ppm | 70 | 35 | [20] |
| Sb-doped WO$_3$ | No | NO$_2$ | 10 ppm | 20 | 51 | [33] |
| WO$_3$ nanorod | No | NO2 | 1 ppm | 50 | 29 | [32] |
| Flower like WO$_3$ | No | NO$_2$ | 200 ppb | 120 | 36 | [31] |
| WO$_3$ | Yes, 150 °C | NH$_3$ | 100 ppm | 25 | 1.02 | This work |
| WO$_3$ | Yes, 150 °C | NO$_2$ | 5 ppm | 25 | 1.01 | This work |

### 3.4. Temperature dependent resistance

Figure 6a depicts the variation in resistance of the samples S1 and S2 as a function of temperature in the range of 30 – 300 $^0$C. Here, the transport measurements were performed on the same samples that were used for NH$_3$ and NO$_2$ detection and also, we note here that the transport properties did not vary before and after the sensor measurements. The resistance of these sensors decreases with increase in temperature and behaves as a typical semiconducting nature. Also, the resistance of these sensors is decreased by more than one order when the temperature is raised from 30 to 300 $^0$C. Here, the high temperature resistance follows thermally activated behavior as shown in Fig. 6b and hence, it is fitted to Arrhenius equation as given below.

$$\rho(T) = \rho(T_0) exp\left(\frac{-E_a}{kT}\right) \dots\dots\dots (2)$$

here $\rho(T_0)$ is a constant, $E_a$ is the activation energy and k is the Boltzman constant. Using the equation 2, the estimated activation energy is found to be ~ 5 and 3 meV for the samples S1 and S2 respectively, in the low temperature regime below 130 $^0$C and it is 157 and 47 meV in the high temperature regime, 170 - 300 $^0$C. The estimated activation energy for both the samples is found to be much smaller as compared to the reported values of 190 – 420 meV in which the activation energy decreases if the $V_O$ concentration increases in $WO_{3-x}$ [34]. The low activation energy also testifies the higher $V_O$ concentration in our $WO_{3-x}$ nanostructures and it manifests as high conductivity in these nanostructures [34]. We note here that the lower resistance of ~ 8.5 kohms of S2 at 25 $^0$C cannot be directly correlated with higher $V_O$ concentration as compared to S1 since these two samples have different thickness and morphology with interconnected 3D network.

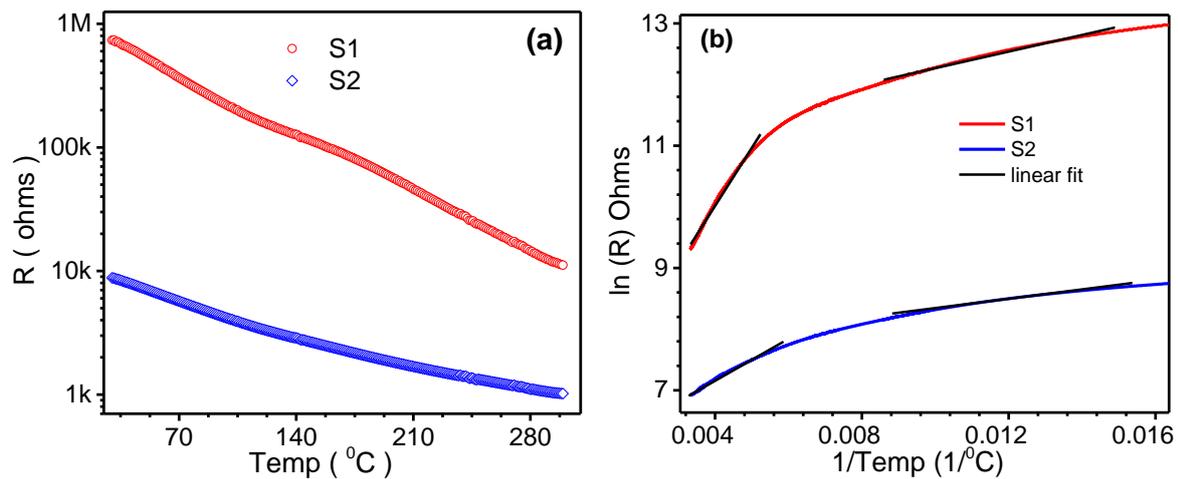

Fig. 6. (a) The temperature dependent electrical resistance of as-prepared $WO_{3-x}$ nanostructures with petal-like structure (S1) and lamella-like structure (S2) and (b) the corresponding Arrhenius plot, 1/T vs ln(R), for the samples, S1 and S2.

After the first heating and cooling cycle, the transport properties of the samples S1_CH$_4$ and S2_CH$_4$ which were annealed under CH$_4$ as described in experimental section were measured. As can be seen from Fig. 7a, the temperature dependent resistance of the sample S1_CH$_4$ behaves like a typical semiconductor when heated upto 300 $^0$C. But, while cooling from 300 $^0$C, the resistance of the sample initially increases gradually down to 100 $^0$C, like a semiconductor and then, it displays an abrupt increase in resistance at 100 $^0$C indicating semiconductor-to-insulator transition (SIT). After this SIT, the resistance becomes nearly constant upon further cooling down to 80 $^0$C. The nature of the abrupt change in resistance signifies the first order transition and the resistance ratio at SIT ($R_S/R_I$) is found to be ~ 35. On the other hand, the simultaneously measured resistance of the sample S2_CH$_4$ varies like a typical semiconductor while heating and cooling cycles, as shown in Fig. 7d.

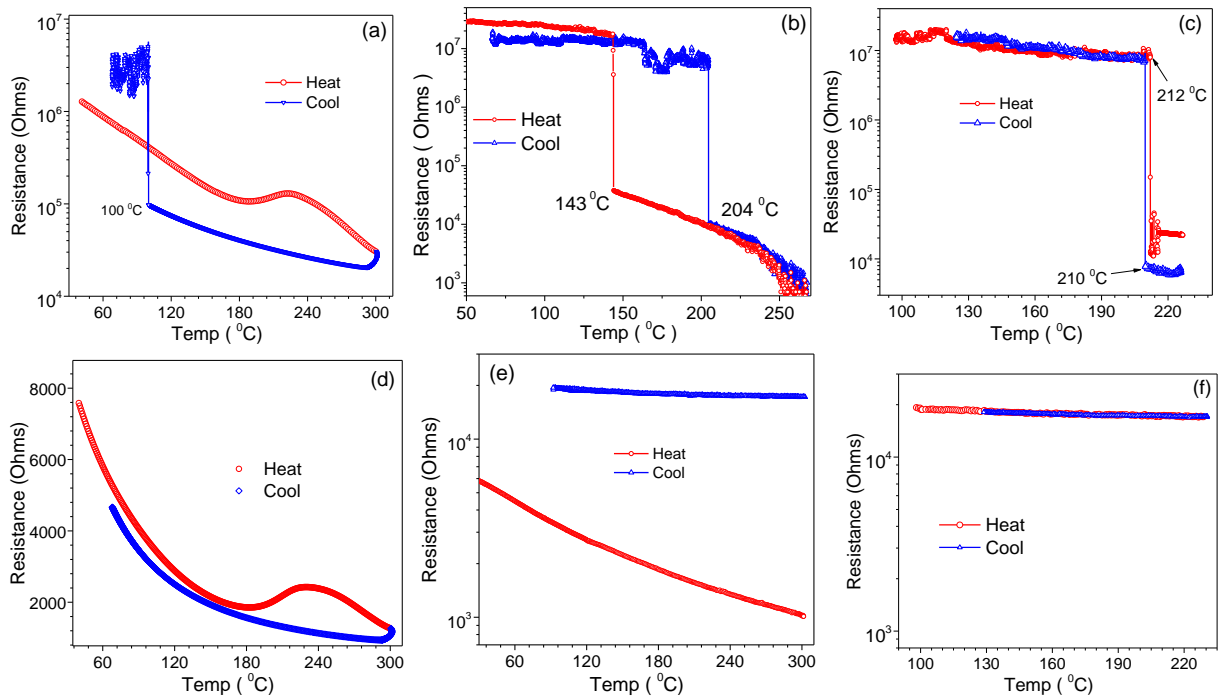

Fig. 7. The variation of resistance with temperature for simultaneously measured WO$_{3-x}$ nanostructures with petal-like (a,b,c) and lamella-like (d,e,f) structure during heating and cooling cycle, (a,d) after CH$_4$ anneal, (b,e) after exposure to ambient air, and (c,f) repeat without breaking vacuum.

After completing the second set of resistance measurements, these samples were further exposed to ambient environment for 12 h and these samples are labelled as S1_CH$_4$_Air and S2_CH$_4$_Air. Subsequently, third set of resistance measurements were performed under vacuum of ~ 8 x 10$^{-3}$ mbar and the results are shown in Figs. 7b and 7e for the samples S1_CH$_4$_Air and S2_CH$_4$_Air, respectively. The sample S1_CH$_4$_Air which retains its insulating character from previous measurement, displays an IST and SIT with abrupt change in resistance at 143 and 204 $^0$C respectively while heating and cooling cycles. Here, the resistance ratios ( $R_I/R_S$ & $R_S/R_I$ ) at transition temperatures of IST and SIT are ~ 500 and 600, respectively. Conversely, the response of the samples S2_CH$_4$_Air, behaves as semiconducting nature while heating upto 300 $^0$C. However, at 300 $^0$C, the resistance of the sample just flipped from semiconducting to insulator instantaneously with ratio $R_S/R_I$ of ~ 20 and then, the sample behaves like an insulator while cooling down to 90 $^0$C. In order to verify the IST and SIT in S1_CH$_4$_Air, the resistance of the sample measured from 90 to 230 $^0$C and subsequently cooled down to 120 $^0$C, as a fourth set of transport measurement. In this time, the vacuum of the chamber was not broken and the samples were maintained in vaccum throughout the measurement. Now, the IST and SIT temperature are shifted to 212 and 210 $^0$C respectively, during heating and cooling cycle, as can be evidenced in Fig. 7c. The resistance ratios ( $R_I/R_S$ & $R_S/R_I$ ) at transition temperatures of IST and SIT are ~ 650 and 950, respectively. Further, the resistance of the simultaneously measured sample S2_CH$_4$_Air remains insulating state in the temperature from 90 to 230 $^0$C, as shown in Fig. 7f.

The MIT in metal oxides is mostly associated with the small polaron conduction mechanism [28,35–38]. The resistivity of the material, $\rho(T)$, with small polaron conduction can be expressed as [35],

$$\frac{\rho(T)}{T} = \rho_\alpha \, exp\left(\frac{E_a}{k_B T}\right) \quad \text{…………………..} \quad (3)$$

where $k_B$ is the Boltzmann constant, $E_a$ is the activation energy, T is the temperature in K and $\rho_\alpha$ is preexponential factor. The plot of $\ln(\frac{\rho(T)}{T})$ vs 1/T will provide the activation energy of the carriers. Fig. 8 displays the plot of $\ln(\frac{\rho(T)}{T})$ vs 1/T for the samples, S1_CH$_4$, S1_CH$_4$_Air, S2_CH$_4$ and S2_CH$_4$_Air corresponding the plots, Fig. 7a, Fig. 7b, Fig. 7d and Fig. 7e respectively, for heating cycle. Further, the activation energy estimated using equation 3 for heating and cooling cycle is presented in Table 2. As observed in Table 2, the activation energy is small whenever the sample undergoes SIT / IST.

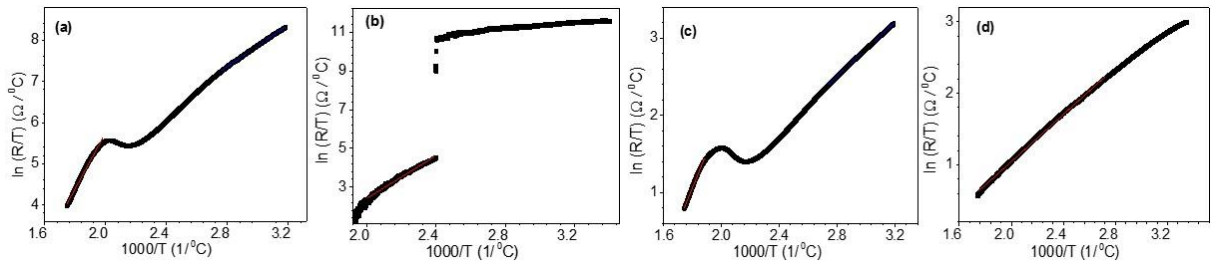

Fig. 8. The plot of $\ln(\frac{\rho(T)}{T})$ vs 1/T for the samples (a) S1_CH$_4$, (b) S1_CH$_4$_Air, (c) S2_CH$_4$ and (d) S2_CH$_4$_Air

Table 2. Activation energy estimated using small polaron conduction mechanism for the $WO_{3-x}$ nanostructures, while heating and cooling cycles

| Sample | S1_CH$_4$ | | S1_CH$_4$_Air | | S2_CH$_4$ | | S2_CH$_4$_Air |
|---|---|---|---|---|---|---|---|
| | High T | Low T | High T | Low T | High T | Low T | 75 – 300 °C |
| $E_a$ (meV) (heating) | 589.6 | 224..0 | 531.0 | 39.6 | 394.7 | 154.2 | 140.0 |
| $E_a$ (meV) (cooling) | 189.7 | - | 474.6 | 65.5 | 155.4 (single slope) | | 48.9 |

## 4. Discussion

Based on the observed results, we briefly summarize a few important points here: EDX analysis clearly reveals a high $V_O$ concentration of 0.57 ± 0.01 for petal-like structures (sample S1) and 0.44 ± 0.01 for lamella-like (sample S2) $WO_{3-x}$ nanostructures. Moreover, Raman spectroscopic studies also confirm the presence of high $V_O$ in both the as-grown $WO_{3-x}$ nanostructures. Moreover, the sensor device behaves as p-type (n-type) conduction at temperatures below 150 $^0$C while it behaves as n-type (p-type) at temperatures above 150 $^0$C for $NH_3$ ($NO_2$) analyte gas. In addition, the petal-like $WO_{3-x}$ nanostructures exhibits much higher sensitivity, especially for $NH_3$, as compared to the lamella-like structure, although both the samples display a similar trend in sensor response. The unusual sensing behavior indicates that more than one charge transport mechanisms contribute to the conduction process in the $WO_{3-x}$ nanostructures.

Before we go into this unusual carrier type conversion, a brief summary of different sensing mechanisms available in the literature is discussed. In general, the change in conductivity occurs due to the charge transfer between the sensor material and the adsorbed molecules that results in surface band bending [39]. Typically, oxygen ions in molecular or atomic form (eg. $O_2^-$, $O^-$ or $O^{2-}$) get adsorbed on the surface of any material. This adsorbed oxygen species can capture electrons from the $WO_3$ structures resulting in a surface depletion layer which would result in increased resistivity. When the sensor material is exposed to reducing gas such as $NH_3$, the electrons captured by the adsorbed oxygen return to the film, leading to decrease in resistivity. The opposite process happens when the material is exposed to an oxidizing gas. This is mostly observed mechanism in chemi-resistive sensor devices.

However, in the present study, we observe a different trend in the response and also, a p- to n- / n- to p- like type conversion as a function of temperature and analyte gas. There are also a few reports available in the literature, similar to our study [11,13,19,40,41]. Also, there are a few models were proposed to explain the anomalous behaviour of the gas sensors [11–13,20,42]. Here, oxygen defects, adsorbed hydroxyl groups, catalytic dissociation of analyte gas at the surface, humidity and crystallite size are crucial factors that affect the sensing properties of a material.

We interpret the unusual sensing behavior of $WO_{3-x}$ nanostructures with characteristic surface band bending as shown in Fig. 9. Since the $WO_{3-x}$ nanostructures have large number of $V_O$, the surface is highly reactive. Hence, the adsorbed moisture and oxygen molecules become $OH^-$, $O^{2-}$, and $O^-$ species by absorbing electron from $WO_{3-x}$ surface. This makes the surface of $WO_{3-x}$ to be depleted with electrons which results in the surface upward band bending with a barrier height of $\phi_{sp1}$. When the oxygen vacancies are very high, the surface of $WO_{3-x}$ nanostructures becomes p-type with inversion layer having thickness of $\delta d_1$ [25]. Fig. 9a displays the surface band bending with p-type inversion layer for $WO_{3-x}$ nanostructures with high $V_O$ concentration at room temperature when exposed to ambient air. When $NH_3$ interacts with surface, it injects more electrons into the surface. Consequently, at low temperatures below 100 $^oC$, the injected electrons compensate the holes near the surface and hence, the surface p-type inversion layer disappears and the surface becomes depletion of electrons with decrease in surface barrier potential height ($\phi_{sp2} < \phi_{sp1}$), as displaced in Fig. 9b. In addition, the concentration of adsorbed molecules such as $OH^-$, $O^{2-}$, and $O^-$ species decreases with increase in temperature ~ 150 $^oC$ and hence, the barrier height decreases further ($\phi_{sp3} < \phi_{sp2}$), as depicted in Fig. 9c. The reduction in barrier height is also reflected in the decrease in positive response, which decreases from ~ 4.5 to 2 % with increase in temperature from 25 $^0C$ to 150 $^0C$. Moreover, the surface barrier potential tends to vanish due to excess electrons by $NH_3$ molecules and hence, the device exhibits a mixed response from surface p-type to bulk n-type conduction at about 150 $^oC$. Above 150 $^oC$, on exposure to $NH_3$, the surface becomes accumulation of excess electrons which results in the decrease of resistance (Fig. 9d).

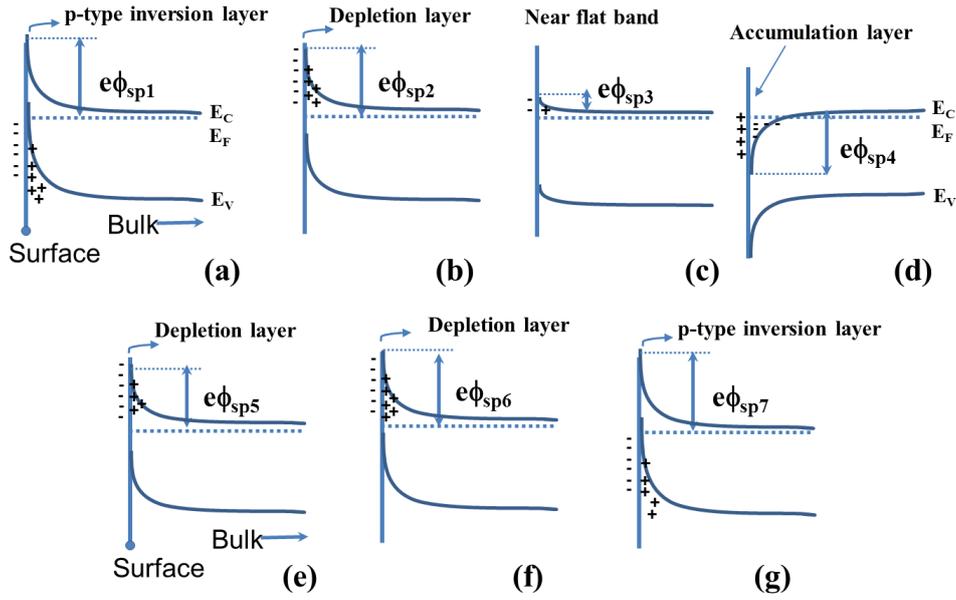

Fig. 9. Energy band diagram of n-type $WO_3$ semiconductor with surface band bending after adsorption of acceptor- and donor-type molecules. (a) Inversion layer after adsorption of atmospheric $O_2$ and $H_2O$ molecules at room temperature, (b) Depletion layer after exposure to $NH_3$ at room temperature, (c) Depletion layer with near flat band potential after exposure of $NH_3$ and also, the removal of partial atmospheric oxidizing agents, ~100 - 150 °C, (d) Accumulation layer at surface after exposure of $NH_3$ at 200 and 300 °C. (e) Depletion layer after exposure to $NO_2$ at room temperature, (f) Depletion layer with Fermi level is close to top of valence band, after exposure of $NO_2$, ~ 100 - 150 °C, (g) Inversion layer after exposure to $NO_2$ at higher temperatures at 200 and 300 °C.

For $NO_2$ detection, the resistance of the samples increases upon exposure to $NO_2$, contrary to the expected decrease at room temperature, as the surface behaves as a p-type semiconductor with an inversion layer. A careful analysis of the literature suggests a possible explanation for this unexpected behavior [12,20,42]. Ruhland et al. [42] demonstrated that $NO_2$ undergoes a complex interaction on the metal oxide surface, leading to its catalytic reduction to NO at low temperatures, as follows.

$$NO_{2\,(gas)} + e^- \rightarrow NO_2^-{}_{(surf)} \quad \ldots\ldots\ldots\ldots\ldots\ldots (4)$$

$$NO_2^-{}_{(surf)} \rightarrow NO_{(gas)} + O^-{}_{(surf)} \quad \ldots\ldots\ldots\ldots\ldots (5)$$

The NO produced at the surface further interacts with $WO_{3-x}$, donating electrons to the surface and leading to electron depletion at room temperature (Fig. 9e). In addition to the depletion of electrons at surface by NO, $NO_2$ molecules also withdraw electrons from the surface, further increasing depletion width. As a the resistance of the $WO_{3-x}$ sensor device increases at temperatures below 150 ºC (Fig. 9f).

However, as the surface potential barrier height continues to increase due to $NO_2$ molecules, an inversion layer forms, leading to p-type conduction on the surface of the $WO_3$ nanostructures. Around 150°C, the conduction process is influenced by both surface and bulk carriers, resulting in a nearly equal contribution of n-type and p-type responses in the sensor device (Fig. 5a at 150°C). When the hole density in the inversion layer becomes sufficiently high, the conduction process is dominated by p-type inversion layer, explaining the decrease in resistance at high temperatures (Fig. 9g). In addition, analyte concentration and temperature also influence the carrier type conversion mechanism. Such change over in the carrier type is manifested as a sharp jump immediately on exposure to the analyte gas, as can be seen from Fig. 5a. Although we present a catalytic behavior for the unusual sensor response to $NO_2$ gas, a detailed probe is essential to confirm the actual mechanism.

Despite the formation of depletion and inversion layers do occur on the surface of every metal oxides, the conduction mechanism is mostly dominated by bulk, in majority of the reports available in literature and hence, this unusual sensing behavior is not observed. However, such an unusual sensor response is possible if the conduction process is dominated by surface rather than bulk. Note that when the crystalline size of the sensor materials becomes comparable to Debye length, surface conductivity dominates in the charge transport mechanism [4]. This can happen when the $WO_3$ nanostructures have a large amount of $V_O$ and one of the dimensions is in the range of a few nanometers. In the present study, a large amount of $V_O$ in the $WO_{3-x}$ nanostructures are introduced by controlling oxygen partial pressure during growth. Further, the thickness of the petal-like structures (S1) is smaller as compared to that of lamella-like structure (S2). Hence, the sample S1 shows higher response than that of sample S2.

An important observation in the study is the discontinuity in resistance as a function of temperature leading to SIT or IST while heating or cooling the $WO_{3-x}$ nanostructures. It has been shown in literature that numerous transition metal oxides exhibit MIT in response to external stimuli, such as temperature, electric / magnetic field, light pressure, chemical doping as well as internal factors like point defects and strain. Studies on such MIT have drawn significant attention from researchers due to their intriguing physics, which helps in understanding fundamental mechanisms and also, offers potential applications across various fields. Only a few studies have reported on MIT in $WO_3$ structures under external stimuli. Here, we present a brief overview of the existing literature on these studies. Researchers have reported the IMT in epitaxial $WO_3$ films and nanostructures using electrolyte-gating processes [36,43–46]. Employing electrolyte gating, the resistivity of $WO_3$ was varied by more than 8 orders of magnitude leading to MIT. Here, the IMT / MIT was induced by either electrostatic accumulation / depletion or electrochemical intercalation / decalation processes using different of types of electrolytes. Leng et. al. [46] proposed that the intercalation of H-atoms into $WO_3$ is responsible for the emergence of a highly conductive ground state and also, provided evidence for the formation of a dense polaronic gas. Altendorf et. al. [44] reported the structural phase transition associated with oxygen removal by an applied gate electric field. In the above mentioned literature review, none of the study has shown or predicted the MIT / IMT as a function of temperature.

In this study, the sample S1_CH$_4$ displays a SIT at 100 $^0$C during cooling from 300 down to 60 $^0$C, maintaining an insulating state down to room temperature. Here, it is expected that an additional amount of V$_O$ concentration is generated in $WO_{3-x}$ nanostructures after annealing under CH$_4$. Upon exposure to air for 12 h under ambient conditions, the insulating state of $WO_{3-x}$ transitions to a semiconducting state (IST) at ~ 143 $^0$C during heating to 300 $^0$C. This semiconducting state then reverts to an insulating state (SIT) at 204 $^0$C during cooling down to 30 $^0$C. It is important to note that the prolonged exposure to air and heating under ~ 8 x 10$^{-3}$ mbar vacuum may have reduced V$_O$ concentration in $WO_{3-x}$ nanostructures. In a repeat experiment conducted without disturbing the vacuum, sample S1_CH$_4$_Air exhibits IST and SIT at 212 and 210 $^0$C, respectively, during the heating and cooling processes. These results reveal a significant variation in the transition temperatures, ranging from 100 to 212 $^0$C. This variation is attributed to differences in the V$_O$ concentration within the $WO_{3-x}$ lattice [37,38]. According to Lu et. al. [37], the MIT temperature in $VO_2$ is tunable and decreases with increasing V$_O$. Additionally, the resistance of the insulating state decreases as V$_O$ concentration increases. Thus, the observed SIT at ~ 100 $^0$C suggests a relatively high V$_O$ concentration in sample S1_CH$_4$ after annealing in CH$_4$. Following exposure to air and subsequent heating the sample upto 300 $^0$C in vacuum, sample S1_CH$_4$_Air undergoes slight oxidation, reducing the V$_O$ concentration in $WO_{3-x}$ and hence, increasing the IST / SIT temperatures. Further, the resistance ratio, R$_S$/R$_I$, is found to be ~ 35, 600 and 950 at the corresponding SIT temperature of ~ 100, 204 and 210 $^0$C, respectively. This increasing resistance ratio aligns with reported data on the MIT transition in $VO_2$ as function of oxygen vacancies [37]. Similarly, sample S2_CH$_4$_Air displays a SIT at relatively higher temperature of ~ 300 $^0$C indicating a lower $V_O$ concentration compared to sample S1_CH$_4$ or S1_CH$_4$_Air.

The oxidation state of W in $WO_{3-x}$ lattice can exist in different states, such as $W^{6+}$, $W^{5+}$ and $W^{4+}$, due to the formation of oxygen vacancies. A single oxygen vacancy introduces two electrons in the system. When the $V_O$ concentration in $WO_{3-x}$ exceeds x ~ 0.12, the electronic structure exhibits strong anisotropy depending upon the formation of vacancy along the -W-O-W- chain direction, as predicted by density functional theory [28]. Wang et. al. [28] demonstrated that $WO_3$ shows a metallic state if the $V_O$ formed along x and y directions while it exhibits semiconducting state for the $V_O$ formation along z direction. Along the z- direction, W ions are connected by a vacancy in the oxidation state of -$W^{5+}$ - $V_O$ (0 e$^-$) -$W^{5+}$- with no trapped electrons at the vacancy site [28]. As shown in Fig.2, Raman spectroscopy provides clear evidence for $W^{5+}$-$W^{5+}$ stretching and O-$W^{5+}$-O bending modes which align with the model predicted by Wang et. al. [28].

Typically, the interaction between electron and phonon is weak, so lattice distortion from point defects do not greatly influence the electrical conduction mechanism. However, in strongly correlated systems like $WO_3$, the strong electron-phonon interactions can localize electrons at lattice distortion, leading to the formation of polaron. When small polarons gain energy exceeding the activation barrier, the polarons are delacalized and the system moves from insulator into either metallic or semiconducting state. In this study, the SIT / IST transitions observed as a function of temperature are attributed to the small polarons conduction mechanism in oxygen deficient $WO_{3-x}$ nanostructures. The polarons surmount activation barrier by absorbing thermal energy, facilitating these electronic transitions. The observed SIT/IST in $WO_{3-x}$ nanostructures as a function of temperature represents the first report of its kind. Here, we attempt to qualitatively describe the SIT/IST in $WO_{3-x}$ using existing theoretical predictions for $WO_3$ and other transition metal oxides. However, a more detailed investigation is required to fully understand the underlying mechanisms of SIT/IST in $WO_{3-x}$ nanostructures.

## 5. Conclusions

Highly oxygen deficient $WO_{3-x}$ nanostructures with petal-like and lamella-like surface morphology were successfully grown directly onto interdigitated electrodes using hot filament chemical vapor deposition (HFCVD). The oxygen vacancy concentration (x) in $WO_{3-x}$ is ~ 0.57 ± 0.01 for petal-like structures and 0.44 ± 0.01 for lamella-like structures, as confirmed by energy dispersive X-ray analysis. Raman spectroscopy further confirms the presence of high oxygen vacancies by exhibiting several low-energy phonon modes. The $WO_{3-x}$ nanostructures exhibited unusual sensor response for both $NH_3$ and $NO_2$ analytes, even at room temperature. Remarkably, the sensing response for both reducing and oxidizing gases was identical under same measurement conditions. This phenomenon is attributed to the carrier type conversion between n-type and p-type behavior upon exposure to analyte gases. The mechanism behind this behavior is explained using the surface band bending phenomenon. Additionally, $WO_{3-x}$ nanostructures exhibit an insulator-to-semiconductor or semiconductor-to-insulator transition (SIT) in the temperature range of 100 - 212 $^0$C depending on the oxygen vacancy ($V_O$) concentration, during the temperature dependent heating and cooling cycles. The transition temperature decreased with increasing $V_O$ concentration in the $WO_{3-x}$ nanostructures. The abnormal gas sensing behavior and SIT are correlated with the presence of $V_O$ in these materials. The high $V_O$ and highly porous nanostructures, with thickness of just a few nanometers, are suggested to play critical roles in the observed unusual sensor responses and the SIT transitions. The key significance of this work is the first observation of SIT as a function of temperature in $WO_{3-x}$ nanostructures. Investigating the physical properties of such highly oxygen-deficient $WO_{3-x}$ nanostructures is particularly intriguing. HFCVD is an effective technique for preparing porous, three dimensional $WO_3$ nanostructures with high $V_O$ concentration. These highly oxygen-deficient $WO_{3-x}$ nanostructures show great promise for applications in gas sensors, catalysts, photochromic and electronic devices.

## CRediT authorship contribution statement

**K. Ganesan:** Investigation, Methodology, Data curation, Formal analysis, Writing – original draft, Writing – review & editing.

**P.K. Ajikumar :** Investigation, Methodology, Data curation, Writing – original draft, Writing – review & editing.

**Declaration of competing interest**
The authors declare that they have no known competing financial interests or personal relationships that could have appeared to influence the work reported in this paper.

**Data availability**
Data will be made available on request.